\documentclass[onecolumn]{aa}
\usepackage{graphicx}
\usepackage{txfonts}

\begin{document}

\title{Tidal interactions of close-in extrasolar planets: the OGLE cases}

\author{M. P\"atzold\inst{1} \and L. Carone\inst{1} \and H. Rauer\inst{2}}

\institute{Institute for Geophysics and Meteorology, University of Colgne, Albertus-Magnus-Platz, D-50923 Cologne \\
          \email{paetzold@geo.uni-koeln.de}
           \and Institute of Planetary Research, Rutherfordstra\ss e 2, D-12489 Berlin
           \email{heike.rauer@dlr.de}}

\offprints{M. P\"atzold}

\date{Received 6 February 2004 / Accepted 19 July 2004}

\abstract{
Close-in extrasolar planets experience extreme tidal interactions with their 
host stars. This may lead to a reduction of the planetary orbit and a spin-up 
of stellar rotation. Tidal interactions have been computed for a number of 
extrasolar planets in circular orbits within 0.06 AU, namely for OGLE-TR-56 b. 
We compare our range of the tidal dissipation value with two dissipation 
models from Sasselov (\cite{sasselov}) and conclude that our choices are 
equivalent to these models. However, applied to the planet OGLE-TR-56 b, we 
find in contrast to Sasselov (\cite{sasselov}) that this planet  will 
spiral-in toward the host star in a few billion years. We show that the 
average and maximum value of our range of dissipation are equivalent to the 
linear and quadratic dissipation models of Sasselov (\cite{sasselov}). Due 
to limitations in the observational techniques, we do not see a 
possibility to distinguish between the two dissipation models as outlined by 
Sasselov (\cite{sasselov}). OGLE-TR-56 b may therefore not serve as a test 
case for dissipation models. The probable existence of OGLE-TR-3 b at 
0.02 AU and the discovery of OGLE-TR-113 b at 0.023 AU and OGLE-TR-132 b at 
0.03 AU may also counter Sasselovs (\cite{sasselov}) assumption of a pile-up 
stopping boundary at 0.04 AU. 

}
\maketitle

\section{Introduction}

Since the discovery of the companion of 51 Pegasi (Mayor \& Queloz \cite{mayor}), many more extrasolar planets orbiting sun-like 
stars of spectral types F, G and K have been discovered in the mass range from 
0.16 Jupiter masses ($M_\mathrm{J}$) to 15 $M_\mathrm{J}$. The total number of 
known extrasolar planets is now 122 (June 2004), including thirteen
 planetary systems with two or more planets. Only the minimum planetary mass 
\mbox{$\tilde{M}_\mathrm{P}=M_\mathrm{P} \sin i$} can be derived with the 
radial velocity method since the inclination $\mathrm{angle}\,i$ of the 
planetary orbital plane cannot be determined unambiguously. The elements of 
these planetary orbits are determined from the analysis of the variation of 
radial velocities of the central star. From the precise determination of the 
radial velocity variation period, interpreted as the extrasolar planet's 
orbital period, and an estimate of the stellar mass, the semi major axis can 
be derived with the aid of Kepler's third law. Surprisingly, planets were found
from very close distances to their host stars to distances as far as 4 AU (the 
outer known planet of 55 Cnc). So far, only one planet has allowed us to derive
its true mass $M_\mathrm{P}$ (and size and therefore density) from its transit 
across the stellar disk of HD209458 in combination with radial velocity 
measurements (Charbonneau et al. \cite{charbon}; Jha et al. \cite{jha};
Henry et al. \cite{henry}). In addition, upper limits of true masses exist 
for few other planets, but are not verified.

The closest extrasolar planet discovered so far, OGLE-TR-56 b  which is also a 
transiting planet, was reported by Konacki et al. (\cite{konackia}) from the 
OGLE survey (Udalski et al. \cite{udalsky}). Improved orbit and planetary 
parameters have been determined by Torres et al. (\cite{torres}). 
Sasselov (\cite{sasselov}) speculates on the stability of the orbit and 
atmosphere. There are further possible transit candidates from the OGLE survey:
OGLE-TR-113 b (Konacki et al. \cite{konackic}, Bouchy et al. \cite{bouchy}) and
OGLE-TR-3 b (Dreizler et al. \cite{dreizler}) which (if the existence of the 
latter is verified) are both as close as OGLE-TR-56 b, OGLE-TR-10 b and 
OGLE-TR-58 b reported by Konacki et al. (\cite{konackib}) and OGLE-TR-132 b 
reported by Bouchy et al. (\cite{bouchy}). The close vicinity of OGLE-TR-56 b 
and OGLE-TR-3 b to their respective host star results in the orbit being 
perturbed by tidal forces exchanged between the star and the planet. We shall 
apply the equations described by P\"atzold \& Rauer (\cite{paetzold}) to these 
OGLE cases which are exact because of the circular nature of their orbits 
($e = 0$). Other planets with circular orbits but which are farther away will 
be included in this work for comparison. We discuss the coupled effects of 
orbital decay and spin-up of the stellar rotation as a function of time in the 
context of the speculations given in Sasselov (\cite{sasselov}).
%-------------------------------------------------------------------
\begin{flushleft}
 \begin{table}[h]
   \caption[]{Parameters of close-in extrasolar giant planets and their host 
              stars with circular orbits with semi major axis $< 0.06 AU$ (as 
of {\bfseries June 2004)}}
    \label{parameter}
      \begin{tabular}{lll*{5}c}	
	\hline\hline
	\bfseries Parameter & & &\bfseries OGLE-TR 56 &\bfseries OGLE-TR 3
        &\bfseries OGLE-TR-113 &\bfseries OGLE-TR 10 &\bfseries  OGLE-TR 132\\
        \hline
        Spectral type & & & G0 & G0 & K & G0 & F\\
        Planetary mass & [$M_\mathrm{J}$] & & 1.450 & 0.500 & 1.35 & 0.700 & 1.010\\
        Semi major axis &[AU]& & 0.023 & 0.023 & 0.023 & 0.042 & 0.0306\\
	revolution period &[days]& & 1.212 & 1.190 & 1.43 & 3.101 & 1.690\\
	Orbit excentricity & & & 0.000 & 0.000 & 0.000 & 0.000 & 0.000\\
	Stellar rotation &[days]& & 18.399 & $27.757^{\mathrm{a}}$ & $27.757^{\mathrm{a}}$ & $27.757^{\mathrm{a}}$ & $27.757^{\mathrm{a}}$\\
        Stellar mass &[$M_{\sun}$]& & $1.040^{\mathrm{b}}$ & $1.000^{\mathrm{b}}$ 
                             & $0.770^{\mathrm{b}}$
                             & $1.000^{\mathrm{b}}$ & $1.340^{\mathrm{b}}$\\
	Stellar radius &[$R_{\sun}$]& & $1.100^{\mathrm{b}}$ & $1.106^{\mathrm{b}}$
                               & $0.822^{\mathrm{b}}$
                               & $1.106^{\mathrm{b}}$ & $1.489{\mathrm{b}}$ 
                              \\\hline
        & &average & 1.230 & 0.870 & $1.080^{\mathrm{d}}$ & $1.300^{\mathrm{d}}$ & $1.150^{\mathrm{d}}$\\
	Planetary $\mathrm{radius}^{\mathrm{c}}$&[$R_\mathrm{J}$] 
        &maximum& 1.972 & 1.383 & & & \\
         &&minimum& 0.861 & 0.604 & &  & \\\hline
	 Roche zone &[$R_{\sun}$]&& 2.706 & 2.721 & 2.022 & 2.721 & 3.663\\
         Stellar $J_2$ &[$10^{-6}$]&& -2.749 & -1.277 & -0.681 & -1.277 & -2.325 
                                   \\
         $I_*$& & & 0.07 & 0.07 & 0.07 & 0.07 &  0.07\\
         $k_{2*}$& & & 0.168 & 0.168 & 0.168 & 0.168 & 0.168\\
	 System property factor &  [$ 10^{-3}M_{\sun}^{\frac {1}{2}}R_{\sun}^5$] & & 2.290 & 0.827 & 0.577 & 1.158 & 6.386\\
        \hline
      \end{tabular}

\bigskip
\begin{tabular}{lll*{5}c}	
	\hline
	\bfseries Parameter & & &\bfseries  OGLE-TR 58 &\bfseries BD-10 3166 &\bfseries HD 209458 &\bfseries HD 76700
        &\bfseries HD 49674\\
        \hline
        Spectral type & & & G0 & G4 & G0 & G6 & G5\\
        Planetary mass & [$M_\mathrm{J}$] & & 1.600 & 0.480 & 0.690 & 0.197 & 0.120\\
        Semi major axis &[AU] & & 0.052 & 0.046 & 0.045 & 0.049  & 0.057\\
	revolution period &[days] & & 4.345 & 3.487 & 3.525 &  3.971 & 4.948\\
	Orbit excentricity & & & 0.000 & 0.000 & 0.000 & 0.000 & 0.000 \\
	Stellar rotation &[days]&  & $27.757^{\mathrm{a}}$ & $27.757^{\mathrm{a}}$ & 15.700 
                         & $27.757^{\mathrm{a}}$ & $27.757^{\mathrm{a}}$\\
        Stellar mass &[$M_*{\sun}$]&  & $0.990^{\mathrm{b}}$ & $1.100^{\mathrm{b}}$ & 
                             $1.050^{\mathrm{b}}$ & $1.000^{\mathrm{b}}$ 
                             & $1.000^{\mathrm{b}}$\\
	Stellar radius &[$R_{\sun}$]& & $1.106^{\mathrm{b}}$ & $1.006^{\mathrm{b}}$ & 
                                $1.200^{\mathrm{b}}$ & $0.956^{\mathrm{b}}$ 
                               & $0.981^{\mathrm{b}}$\\\hline
        & &average  & $1.600^{\mathrm{d}}$  & 0.850 & 0.970 & 0.632 & 0.535\\
	Planetary $\mathrm{radius}^{\mathrm{c}}$ &[$R_\mathrm{J}$] 
        & maximum & & 1.349 & 1.540 & 1.003 & 0.850\\
        & &minimum & & 0.589 & 0.673 & 0.438 & 0.371\\\hline
	Roche zone &[$R_{\sun}$]& & 2.721 & 2.475 & 2.952 &  2.352 &  2.414\\
        Stellar $J_2$ &[$10^{-6}$]& & -1.290 & -0.874 & -4.855 & -0.825 & -0.892\\
        $I_*$& & & 0.07 & 0.07 & 0.07 & 0.07 & 0.07\\
        $k_{2*}$& & & 0.168 & 0.168 & 0.168 & 0.168 & 0.168\\
	System property factor & [$ 10^{-3}M_{\sun}^{\frac {1}{2}}R_{\sun}^5$] &  & 2.661 & 0.472 & 1.676 & 0.157 & 0.109\\
        \hline
      \end{tabular}
\begin{list}{}{}
\item[$^{\mathrm{a}}$] If the true stellar rotation period is not known, the 
value was set, without loss of generality, to the sun's rotation period.
\item[$^{\mathrm{b}}$] Derived from Aller et al. (\cite{aller}).
\item[$^{\mathrm{c}}$] If the planetary radius is not known, then 
upper and lower limits can be derived by setting the planetary bulk density to 
$250\,\frac{\mathrm{kg}}{\mathrm{m}^3}$ (Henry et al. \cite{henry}; 
Jha et al. \cite{jha}) and $3000\,\frac{\mathrm{kg}}{\mathrm{m}^3}$ 
(Cameron et al. \cite{cameron}), respectively. As an average, the density of $1000\,\frac{\mathrm{kg}}{\mathrm{m}^3}$ was assumed.
\item[$^{\mathrm{d}}$] This is the true radius as derived from transit 
observations ( Charbonneau et al. \cite{charbon}, Konacki et al. \cite{konackib}).
\end{list}
 \end{table}
\end{flushleft}
%-------------------------------------------------------------------
\section{Tidal interaction}
\label{tidal}

Close-in extrasolar planets experience strong tidal interactions with 
their central star. If the planet's orbital period $P$ is smaller than the 
star's rotation period $P_*$, tidal friction will lead to a spin-up of the 
star and, due to the conservation of momentum, will also lead to a decrease of 
the semi major axis of the planet's orbit.The parameters of the planets and 
the stars considered in this work are listed in Table \ref{parameter}. 
Only planets with circular orbits within 0.06 AU are used in this study. A more
detailed study for elliptical orbits in general is in preparation.

The rotation periods of only two stars (OGLE-TR-56 via \mbox{$v \sin i$} and 
HD209458) in Table \ref{parameter} are known. Due to the 
small revolution periods of the planets, all other stars must be fast rotators 
(\mbox{$P_* < 3 \mathrm{days}$}) and the condition \mbox{$P_* > P$} for the 
decrease of the orbit does not hold and is considered as not very 
probable. The stellar rotation periods are set to the solar rotation period of 
27 days if these parameters are not explicitly known or 
estimated otherwise. Following Murray \& Dermott (\cite{murray}), 
Goldreich \& Soter (\cite{goldreich_a}), 
Goldreich \& Nicholson (\cite{goldreich_b}) and Zahn (\cite{zahn}), we derive 
for the change in stellar rotation and the change in orbit radius
\begin{eqnarray}
\frac{d\Omega_*}{dt}& =&-\mathrm{sign}\,(\Omega_* - n)\frac{3k_{2*}}{2C_*Q_*}
                       \left(\frac{\tilde{M}^{2}_{\mathrm{P}}}{M_*} \right)
                       \left(\frac{R^{5}_*}{a^{3}}\right)\frac{GM_*}{a^3}
\label{rot}\\
\frac{da}{dt} &=& \mathrm{sign}\,(\Omega_*-n)\frac{3k_{2*}}{Q_*}
                \left(\frac{\tilde{M}_{\mathrm{P}}}{M_*} \right)
                 \left(\frac{R_*}{a}\right)^5 a \sqrt{\frac{GM_*}{a^3}}
\label{a}
\end{eqnarray}
where $\Omega_*$ and $n$ are the stellar rotation and planetary revolution, 
respectively, $k_{2*}$ is the stellar Love number, $C_*$ is the stellar moment 
of inertia, $Q_*$ is the stellar dissipation factor which includes tidal 
friction, $M_*$ and $R_*$ are the stellar mass and radius, respectively, $a$ is
 the semi major axis of the planetary orbit and $G$ is the gravitational 
constant. The relations (\ref{rot}) and (\ref{a}) are valid for a circular 
orbit \mbox{$(e = 0)$} which is assumed from the observations for the OGLE 
candidates. We used the stellar moment of inertia $C_*$ which is usually 
expressed in (\ref{rot}) as \mbox{$C_* = I_*M_*R^2_*$} where 
$I_*$ is the normalized stellar moment of inertia. The mass distribution of 
the standard solar model, as well as the GONG model, yield an 
\mbox{$I_* = 0.07$} (Bahcall et al. \cite{bahcall}, 
Christensen-Dalsgaard et al. \cite{christ}). Both the stellar Love number 
$k_{2*}$ and the tidal energy dissipation factor $Q_*$ are only poorly known. 
We use \mbox{$10^6\!<\!Q_*\!<\!3\cdot 10^7$}, and 
\mbox{$0.02\leq k_{2*}\leq 0.17$} which results in 
\mbox{$5\cdot 10^6\leq \frac{Q_*}{k_{2*}}\leq 1.5\cdot 10^9$} 
(P\"atzold \& Rauer \cite{paetzold}) for these two combined parameters, with 
\mbox{$\frac{Q_*}{k_{2*}} = 1.2\cdot 10^8$} as an average. This is in agreement
with the range of energy dissipation for main sequence stars used in the 
literature (Lin et al. \cite{lin}; Trilling et al. \cite{trilling}; 
Bodenheimer et al. \cite{bodenheimer}). Fixing the Love number at 
\mbox{$k_{2*}=0.17$} (P\"atzold \& Rauer \cite{paetzold}), the average 
dissipation factor is in agreement with the value found by  
P\"atzold \& Rauer (\cite{paetzold}) for the F-stars;
a Love number of \mbox{$k_{2*}=0.02$} yields \mbox{$Q_*=2.4\cdot 10^6$}.

The stellar radius is another important input parameter for the relations 
(\ref{rot}) and  (\ref{a}). We derive these values from calibration curves 
based on stellar evolution models (Aller et al. \cite{aller}). 
Integration of (\ref{rot}) yields for the stellar rotation
\begin{equation}
\Omega_*(t)=-\frac{\tilde{M}_\mathrm{P}\sqrt{GM_*}}{C_*}
             \left \{ \left [ a_{0}^{\frac{13}{2}} + \mathrm{sign}\,(\Omega_*-n)
             \frac{13}{2}\frac{3k_{2*}}{Q_*}\frac{\tilde{M}_{\mathrm{P}}}{M_*}
             R^5_*\sqrt{GM_*}t\right ]^{1/13} - 
             \sqrt{a_0}\right \}+\Omega_{*0}
             \label{rot_t}
\end{equation}
and for the semi major axis
\begin{equation}
a(t)=\left[a_0^{\frac{13}{2}}+\mathrm{sign}\,(\Omega-n)\frac{13}{2}
     \frac{3k_{2*}}{Q_*}\frac{\tilde{M}_\mathrm{P}}{M_*}R^5_*\sqrt{GM_*}t\right]^
     \frac{2}{13}
     \label{a_t}
\end{equation}
The time scale for spiralling inward to the stellar Roche limit 
\mbox{$a_\mathrm{roche}=2.46\,R_*$} (Chandrasekhar \cite{chandra}) from the 
present semi major axis $a_0$ is (as derived from (\ref{a_t}))
\begin{equation}
\tau_{a*} = \frac{\frac{2}{13}\left[a^{\frac{13}{2}}_0 - 
            a^{\frac{13}{2}}_\mathrm{roche}\right]}{3\frac{k_{2*}}{Q_*}
            \frac{\tilde{M_\mathrm{P}}}{M_*}R^5_*\sqrt{GM_*}}
            \label{tau_a}
\end{equation}
Here and in all following equations we have replaced the true planetary mass 
\mbox{$M_\mathrm{P}$} with the minimum mass 
\mbox{$\tilde{M_\mathrm{P}} = M_\mathrm{P}\sin i$}. Aside from the larger 
uncertainties of $Q_*$, the time scale \mbox{$\tau_{a*}$} in (\ref{tau_a}) has 
to be considered as an upper limit because the true planetary mass 
(\mbox{$M_\mathrm{P} \geq \tilde{M_\mathrm{P}}$}) will further decrease this 
value. The time scale $\tau_{a*}$ represents the remaining time for the tidal 
process and not the time since formation of the stellar system.
\begin{figure}
   \centering
   \includegraphics[width=\textwidth]{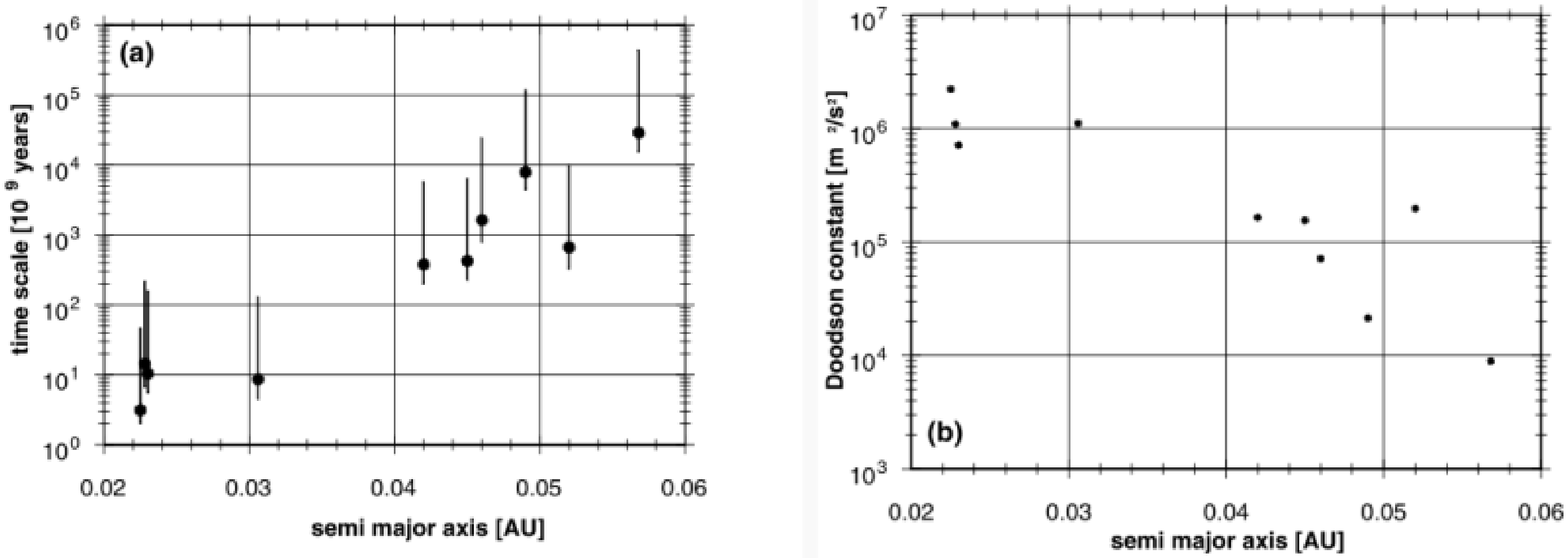}
      \caption{\textbf{a)} The time scales for spiralling into the host stars 
               (according to equation (\ref{tau_a})) of the extrasolar planets 
               located at semi major axis of today. The symbols mark the 
               avarage value of $\frac{Q_*}{k_{2*}}$, the error bar the 
               selected range of $\frac{Q_*}{k_{2*}}$ 
               (see section \ref{tidal}).
	       \textbf{b)} The Doodson constant of extrasolar planets with 
	       $e=0$ for a semi major axis less than 0.06 AU.
              }
         \label{fig_timescale}
   \end{figure}
\section{Discussion}
\label{discuss}

Figure \ref{fig_timescale} (a) shows the time needed to let the planets spiral 
into the central star starting from their orbital position of today. The error 
bars reflect the variation of the parameter \mbox{$\frac{k_{2*}}{Q_*}$}. The 
closest planets, OGLE-TR-56 b and OGLE-TR-3 b, will spiral in within a few 
billion years on average, even much faster considering the lowest boundary of 
\mbox{$\frac{k_{2*}}{Q_*}$.} The upper boundary of \mbox{$\frac{k_{2*}}{Q_*}$} 
is obviously generously chosen so that no influence on the orbit can be seen. 
This orbit may be considered as stable. Possible candidates for strong orbital 
decay can be found by examining the system property factor 
\mbox{$\frac{M_\mathrm{P}}{\sqrt{M_*}}R^5_*$} 
(P\"atzold \& Rauer \cite{paetzold}). Another possibility is the well known 
Doodson constant (Figure \ref{fig_timescale} (b)) which describes the amplitude
of tidal forces exchanged between the two partners:
\begin{equation}
D_\mathrm{P} = \frac{3}{4}\frac{GM_\mathrm{P}}{a^3}R^2_*
               \label{doodson}
\end{equation}
Although our model does include viscosity and friction inside the star 
expressed by the dissipation factor $Q_*$, we do not agree with 
Sasselov (\cite{sasselov}) who considers the existence of OGLE-TR-56 b as 
counterproof of the orbit decay description in P\"atzold \& Rauer 
(\cite{paetzold}). The computed time scale (\ref{tau_a}) is counted from today 
to the time of planetary disruption and not from the time of creation of the 
planetary system. Having in mind the many theories of planet formation, 
migration and tidal movement, it is very difficult to assess how the close-in 
planets finally achieved the current observed positions. Let us compare the 
equation (\ref{a}) used by P\"atzold \& Rauer (\cite{paetzold}) and the 
relation used by Sasselov (\cite{sasselov}), which was given by 
Rasio et al. (\cite{rasio}):
\begin{equation}
  \frac{\dot{a}}{a}=\frac{1}{\tau_{a*}}=\frac{f}{\tau_\mathrm{C}}
                  \frac{M_\mathrm{CZ}}{M_*}\frac{M_\mathrm{P}}{M_*}
                  \left\{1+\frac{M_\mathrm{P}}{M_*}\right\}\frac{R_*}{a}^8
                  \label{adot}
\end{equation}
where 
\mbox{$f = \min \left[1,\left (\frac{1}{n\tau_\mathrm{C}}\right )\right]$} or 
\mbox{$f = \min \left[1,\left (\frac{P}{2\tau_\mathrm{C}}\right )\right]$} 
is the total integrated turbulent viscosity in the quadratic viscosity 
suppression case and the linear suppression case, respectively. 
\mbox{$\tau_\mathrm{C} = 18\, \mathrm{days}$} is the turbulent time scale 
(Sasselov \cite{sasselov}) and \mbox{$M_\mathrm{CZ} = 0.1 M_*$} is the mass of 
the convection zone for a sun-like star as derived from the standard solar 
model. Comparing directly (\ref{a}) with (\ref{adot}), we have:
\[
\begin{array}{lllp{0.8\linewidth}}
     \frac{k_{2*}}{Q_*} & = &  \frac{1}{3}\frac{0.01}{n\tau^3_\mathrm{C}} 
                              \frac{R^3_*}{GM_*} & for the quadratic 
                                                   suppression case and \\
     \frac{k_{2*}}{Q_*} & = & \frac{1}{3}\pi\frac{0.01}{\tau^2_\mathrm{C}}
                             \frac{R^3_*}{GM_*} & for the linear 
                                                   suppression case.
\end{array}
 \]
The former one yields 
\mbox{$\left.\frac{Q_*}{k_{2*}} \right|_\mathrm{quadratic} \approx 2.7 \cdot 10^{10}$} which is comparable to the upper boundary of our chosen range of $Q_*$ 
while the latter one yields 
\mbox{$\left.\frac{Q_*}{k_{2*}} \right|_\mathrm{linear} \approx 9 \cdot 10^7$} 
which is slightly smaller than our chosen average value of $Q_*$.
\begin{figure}
   \centering
   \includegraphics[width=\textwidth]{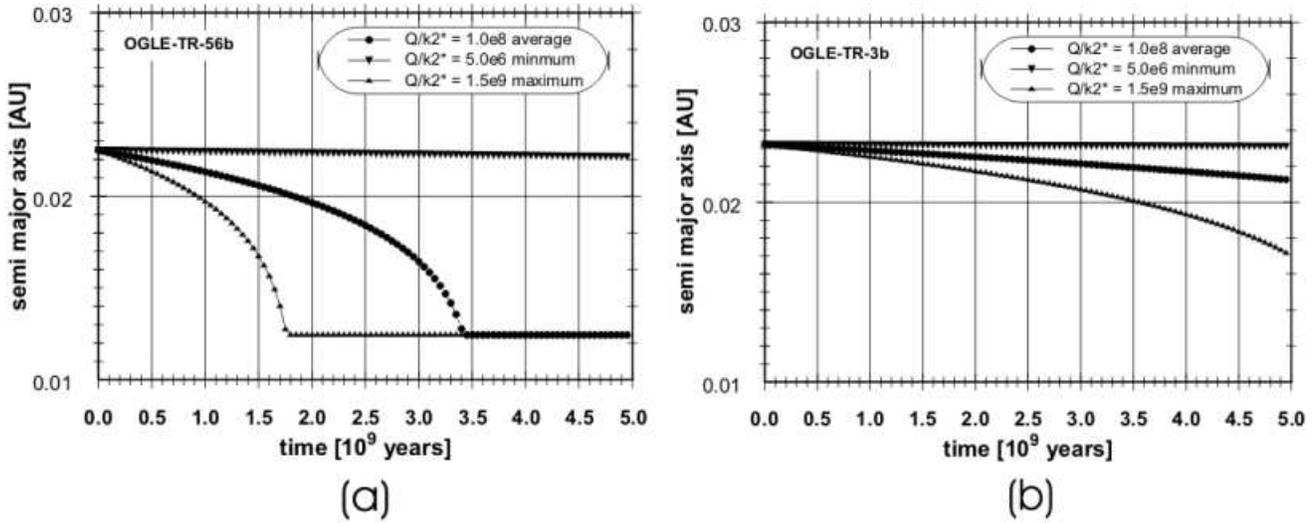}
      \caption{The change in semi major axis of OGLE-TR-56 b \textbf{a)} and 
               OGLE-TR-3 b \textbf{b)} within 5 billion years for different 
	       values $\frac{Q_*}{k_{2*}}$. For values of $\frac{Q_*}{k_{2*}}$ 
               smaller than the avarage, the planet needs less than a few 
               billion years to spiral into the star.
              }
         \label{fig_da}
   \end{figure}
Figure \ref{fig_da} shows the time dependence of the orbital decay of 
OGLE-TR-56 b and OGLE-TR-3 b. The three curves describe the minimal, average 
and maximal value of \mbox{$\frac{Q_*}{k_{2*}}$} as discussed above in 
(\ref{tidal}). For OGLE-TR-56 b, the average value of 
$\frac{Q_*}{k_{2*}} = 10^8$ lets the planet spiral into the star (Roche limit 
of $2.46 R_*$) within a few billion years, the lower value of 
\mbox{$\frac{Q_*}{k_{2*}}$} achieves the same result in about 1.5 billion 
years. If the linear suppression case which considers dissipation is true, the 
planet is doomed, since in that case \mbox{$\frac{Q_*}{k_{2*}}$} is even 
smaller than our chosen average value. No effect within the life time of the 
star is observed for the upper boundary value of \mbox{$\frac{Q_*}{k_{2*}}$} 
which is more or less identical with the quadratic suppression case. 
Figure \ref{fig_da} (b) shows the temporal dependence of OGLE-TR-3 b. Although 
this planet is slightly further away from its central star than OGLE-TR-56 b, 
its mass is lower ($0.5 M_\mathrm{J}$ (Dreizler, priv. Comm.)) and its Doodson 
constant is smaller. The result is that the tidal effects are weaker than for 
OGLE-TR-56 b (Figure \ref{fig_timescale} (b)) and Figure \ref{fig_da} (b)).

\begin{figure}
   \centering
   \includegraphics[width=\textwidth]{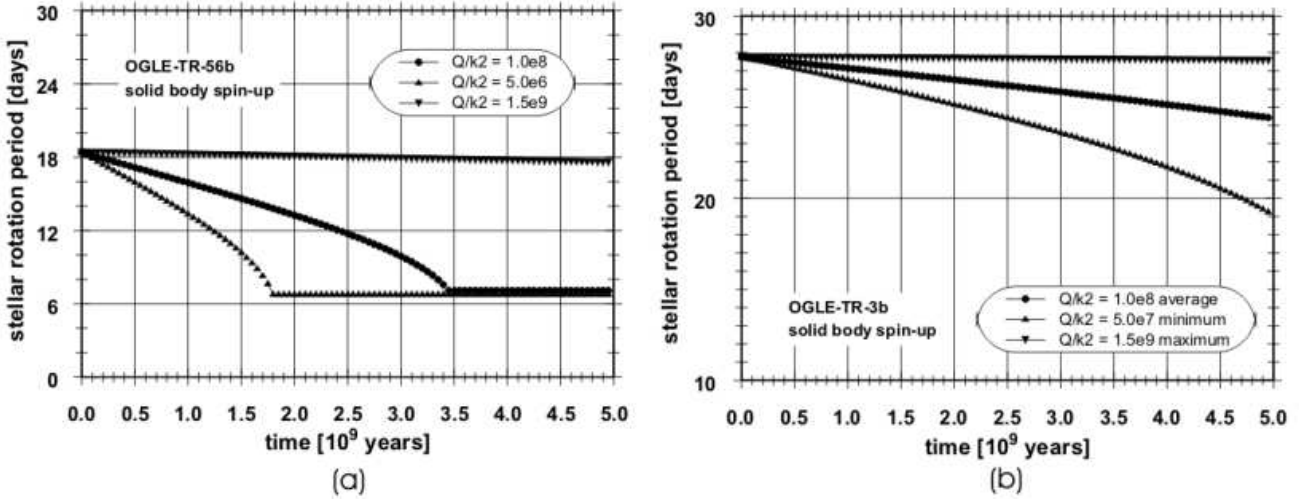}
      \caption{The change in stellar rotation period of the host stars of 
               OGLE-TR-56 b \textbf{(a)} and OGLE-TR-3 b \textbf{(b)} due to 
	       tidal friction.In this case, it was assumed that the stellar 
	       body is rigid. For values of $\frac{Q_*}{k_{2*}}$ smaller than 
	       the avarage, the star is significantly spun up.
              }
         \label{fig_rotsol}
   \end{figure}

Due to the conservation of angular momentum, the star is spun up as the orbit 
decays. Equation (\ref{rot}) describes the stellar spin-up by assuming the star
being a solid body, represented by the introduction of the total moment of 
inertia relative to the rotation axis $C_*$. We neglected friction within the 
star. Figure \ref{fig_rotsol} shows that the spin-up of the entire solid body 
according to (\ref{rot}) leads to a rapid decrease of the rotation period from 
18 days to less than 15 days within the time of 800 Million years needed for 
the planet to spiral-in towards the stellar Roche zone. For the upper boundary 
of $\frac{Q_*}{k_{2*}}$ no significant effect can  be seen.
It may be argued that mass loss of the star due to stellar wind decreases the
angular momentum with increasing stellar age and thus leads to a decrease in 
stellar rotation, counteracting the tidal spin-up. But we assume that this 
effect can safely be neglected since our dataset contains only sun-like stars 
which will not lose a considerable amount of mass during the next four to five 
billion years. However, we did not consider friction within 
the stellar body and effects of the stellar magnetic field which might 
counteract the tidal spin-up of the stellar body.

\begin{figure}
   \centering
   \includegraphics[width=\textwidth]{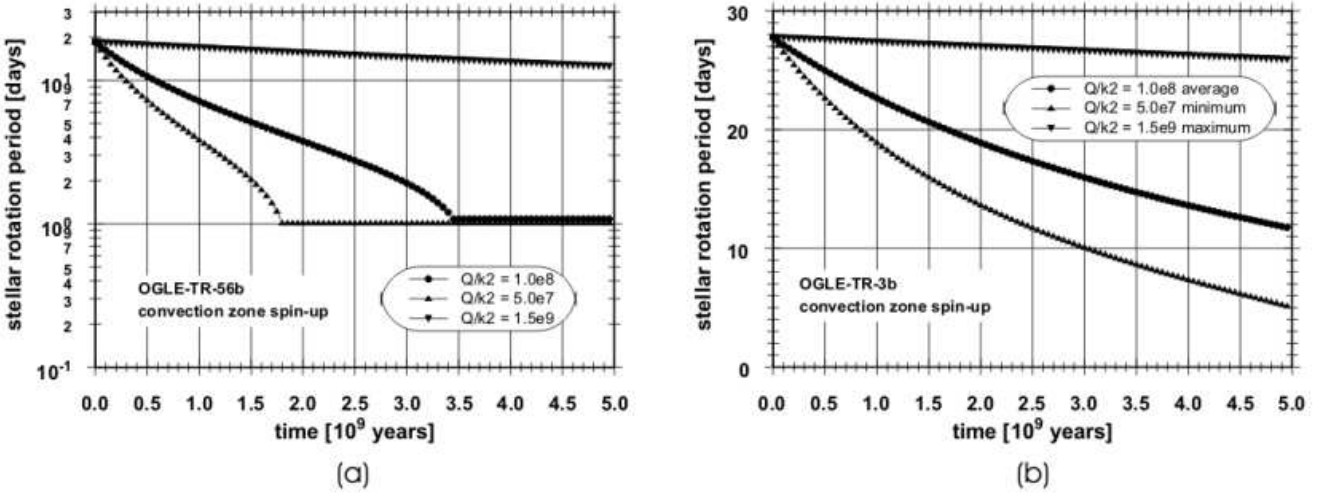}
      \caption{The change in stellar rotation period of the host stars of 
               OGLE-TR-56 b \textbf{a)} and OGLE-TR-3 b \textbf{b)} due to 
	       tidal friction. In this case, it was assumed that the tidal 
	       enery is dissipated in the outer stellar layers. Neglecting 
	       friction between the outer layers and the core, the spin-up of 
	       the outer stellar layer is significant even for the maximum 
	       value of $\frac{Q_*}{k_{2*}}$.
              }
         \label{fig_rotcz}
   \end{figure}

There is discussion that not the entire stellar body is spun up, but only the 
convection zone (Rasio et al. \cite{rasio}, Sasselov \cite{sasselov}). That would 
mean that in (\ref{rot}) the moment of inertia of the entire body needs to be 
replaced by the moment of inertia of the convection cell.
\begin{equation}
\frac{d\Omega_*}{dt} = -\mathrm{sign}\,(\Omega_* - n)
                       \frac{3k_{2*}}{2C_\mathrm{CZ}Q_*}
                       \left(\frac{M^{2}_{\mathrm{P}}}{M_*} \right)
                       \left(\frac{R^{5}_*}{a^{3}}\right)\frac{GM_*}{a^3}
\label{rot_conv}
\end{equation}
where \mbox{$C_\mathrm{CZ} = \frac{2}{5}\sum_{i=r_\mathrm{C}}^{R_*}m_i
                       \frac{r^5_{i+1}-r^5_i}{r^3_{i+1}-r^3_i}M_*R^2_*
                     = I_\mathrm{CZ}M_*R^2_*$ and $I_\mathrm{CZ}$} is the
normalized moment of inertia of the convection zone, $r_\mathrm{c}$ is the 
radius of the inner boundary of the convection zone in stellar radii, and $m_i$
is the mass of the layer between $r_{i+1}$ and $r_i$ in stellar masses. Using 
again the standard solar model, the normalized moment of inertia of the 
convection zone is \mbox{$I_\mathrm{CZ} = 0.01$},\mbox{ $m_{CZ} = 0.01$} and 
\mbox{$r_\mathrm{c} = 0.75$}. Figure \ref{fig_rotcz} shows the change of the 
rotation of the stellar convection zone.
\section{Conclusion}

We do not agree with Sasselov (\cite{sasselov}) that his derived long orbit 
decay time of about 14.3 billion years for the quadratic suppression case is a 
proof of his description of turbulent viscosity dissipation, simply justified 
by the fact that the planet still exists. The derived time scales in 
(\ref{tau_a}) describe the remaining time for orbit decay and not the time 
since the formation of the system. On average, OGLE-TR-56 b will spiral into 
the Roche zone of its respective host star within 3.5 billion years. Because of
its lower mass, OGLE-TR-3 b may be safe, if this planet exists at all. 
Correcting Sasselov (\cite{sasselov}), we have shown that the suppression case
for dissipation in the stellar convective layers is considered by our choice of
$\frac{Q_*}{k_{2*}}$. Assuming the linear suppression case, the dissipation in 
the convection zone is considered as our average value of $\frac{Q_*}{k_{2*}}$.
The quadratic suppression case is covered by our upper boundary value for 
$\frac{Q_*}{k_{2*}}$. Applying this model would yield absolutely no tidal 
effect for any planet at any distance.

Would it be possible to distinguish between these two dissipation models? 
This would mean to measure the orbital decay rate of the planet via the
 transit time across the stellar disk. The velocity in a circular orbit is
\mbox{$v^2=\frac{GM_*}{a}=\frac{R^2_*}{\Delta t^2}$}. Assuming that the orbital
radius $a$ is decaying with the rate at first order (\ref{a}) 
\mbox{$a(t_2)=a(t_1)-\frac{da}{dt}(t_2-t_1)$} it is clear that
\begin{equation}
\Delta t^2(t_2) - \Delta t^2(t_1) = \frac{R^2_*}{GM_*}\frac{da}{dt}(t_2 - t_1)
                                  = \frac{3k_{2*}}{Q_*}
                                    \frac{M_\mathrm{P}}{M_*}
                                    \left(\frac{R_*}{a}\right)^7
                                    \sqrt{\frac{a^3}{GM_*}}(t_2-t_1)
\label{delta_t}
\end{equation}
Thus, the transit times would change by 0.1 seconds in one hundred years or 
by 1 ms per year for the average value of $\frac{Q_*}{k_{2*}}$, and 
would require a timing accuracy of the transit times better than 
0.1 millisecond. With the current observation methods and integration times 
this accuracy is simply not achievable. Therefore, one would not be able to 
judge which of these models describes the convective dissipation and 
the proposal of Sasselov (\cite{sasselov}) to use OGLE-TR-56 b as a test case 
for convection dissipation models cannot hold.

We have also shown that the consequence of orbital decay results in
the spin-up of the stellar rotation. For all selected values of 
$\frac{Q_*}{k_{2*}}$, the convection zone is spun-up significantly. If this is 
true, one should find all stars with close-in extrasolar planets to be fast 
rotators. However, this is not observed. Friction between the convection 
zone and the remaining inner radiation zone and core avoids a fast acceleration
of the convection zone, in particular of the observable outer layers.

We also do not agree with the conception of a 'pile-up' of close-in planets at 
about 0.04 AU due to an unknown stopping mechanism for the inward migration 
(Sasselov \cite{sasselov}). First, this concept is based on an extremely 
limited observation base which does not allow any conclusions. Second, we have 
shown here that the orbital decay of planets closer than 0.04 AU is fast if the
planetary mass is sufficiently large and only in rare cases allows the 
observation of planets within 0.04 AU which are in the process of spiraling 
toward the host star. Third, OGLE-TR-3 b, if it exists at all, is the 
counterproof: due to its low mass the tidal interaction is negligible.

\begin{acknowledgements}
This work was funded by the \emph{Deut\-sche For\-schungs\-an\-stalt f\"ur 
Luft- und Raum\-fahrt, DLR\/}, Bonn, Germany under grant 50 OW 0203.
\end{acknowledgements}

\end{document}